\documentclass[aps,pre,twocolumn,floats,amsmath,amssymb,showpacs,floatfix]%
{revtex4}
\usepackage{graphicx}
\usepackage{dcolumn}

\begin{document}
\title{Clustering of exponentially separating trajectories}
\author{M. Wilkinson$^{(1)}$, B. Mehlig$^{(2)}$,
K. Gustavsson$^{(2)}$, E. Werner$^{(2)}$}
\affiliation{$^{(1)}$Department of Mathematics and Statistics,
The Open University, Walton Hall, Milton Keynes, MK7 6AA, England\\
$^{(2)}$Department of Physics, Gothenburg University, 41296
Gothenburg, Sweden}

\begin{abstract}
It might be expected that trajectories for a dynamical system which has no negative Lyapunov exponent (implying exponential growth of small separations) will not cluster together. However, clustering can occur such that the density $\rho(\Delta x)$ of trajectories within distance $\Delta x$ of a reference trajectory has a power-law divergence, so that $\rho(\Delta x)\sim \Delta x^{-\beta}$ when $\Delta x$ is sufficiently small, for some $0<\beta<1$. We demonstrate this effect using a random map in one dimension. We find no evidence for this effect in the chaotic logistic map, and argue that the effect is harder to observe in deterministic maps.
\end{abstract}

\maketitle

{\sl 1. Introduction}. It is well known that the trajectories of simple dynamical systems can approach fractal sets known as strange attractors \cite{Eck+85,Ott02}. These arise when the largest Lyapunov exponent is positive, implying that trajectories diverge exponentially, but where there is contraction (characterised by negative Lyapunov exponents) in other directions, so that volume elements contract. Here we demonstrate an alternative mechanism for trajectories of a dynamical system to exhibit fractal clustering effects. We demonstrate, explain and analyse how fractal clustering of trajectories may occur in a system which has no negative Lyapunov exponents, so that there is no attractor.

Fractal measures which arise in the study of dynamical systems are typically multifractal, in the sense that their Renyi dimensions $D_q$ (discussed in \cite{Hen+83,Ott02}) are not all equal. Exponential separation in all directions implies that the invariant density must cover a set of finite measure, so that for systems with only positive Lyapunov exponents the box-counting dimension $D_0$ is equal to the space dimension, $d$. Furthermore, the Kaplan-Yorke formula \cite{Kap+79} for the Lyapunov dimension $D_{\rm L}$ gives $D_{\rm L}=d$ if the sum of the Lyapunov exponents is positive. It is believed that $D_{\rm L}=D_1$ \cite{Led+88}, so that we conclude that $D_1=d$ for the systems which we consider. There does not appear to be any constraint which implies $D_2=d$ when all of the Lyapunov exponents are positive. It is, however, hard to conceive of how fractal clustering can occur in a situation where trajectories are separating exponentially in each direction. If a cluster with arbitrarily high density is to form, the trajectories which participate have to \lq beat the odds' by coming closer together throughout a long sequence of iterations of the map, despite the fact that positive Lyapunov exponents imply that separations are more likely to increase than to decrease.

2. {\sl The Lyapunov exponent and correlation dimension}. We will discuss the simplest case which illustrates our conclusion, which is that of chaotic maps in one dimension. The extensions to flows and to higher dimensions are straightforward. We start by reviewing the definitions of the Lyapunov exponent $\lambda$ and of the correlation dimension $D_2$. It is possible to describe the fractal attractor of a deterministic map by considering a measure on the set of points visited by a single trajectory. In this paper, however, we shall consider random as well as deterministic maps, so that the fractal measures upon which trajectories might congregate are not fixed in the coordinate space. We will, therefore, describe clustering in terms of the behaviour of trajectories with different initial conditions, examined at the same \lq time' (that is, iteration number). The initial distribution of the trajectories is a random scatter across the coordinate space, with uniform density.

The infinitesimal separation $\delta x$ of two trajectories is characterised by the Lyapunov exponent, $\lambda$ \cite{Ott02}. For a one-dimensional map generating a sequence of points $x_1,x_2,\ldots,x_i,\ldots$ this is defined by writing
\begin{equation}
\label{eq: 2.1}
\lambda=\lim_{N\to \infty}\frac{1}{N}\left\langle {\rm ln}\frac{\delta x_N}{\delta x_1} \right\rangle
\end{equation}
where the $\delta x_i$ are infinitesimal separations of two trajectories at the $i$th iteration (throughout this paper $\langle X\rangle$ will denote the expectation value of $X$).
If $\lambda <0$, the separation of two very close orbits will approach zero with a probability which approaches unity as the initial separation approaches zero. If $\lambda >0$, the separation of two orbits which are initially very close increases with a probability which approaches one as the initial separation decreases. Given this observation, it would be surprising to see trajectories clustering together if $\lambda >0$. Here we will show that clustering may occur, such that the probability density $\rho$ for the separation of trajectories, $\Delta x$ has a power-law dependence for small values of $\Delta x$, i.e.
\begin{equation}
\label{eq: 2.2}
\rho(\Delta x)\sim \Delta x^{-\beta}
\end{equation}
with $0<\beta < 1$.

The Renyi dimensions $D_q$ are defined by dividing the configuration space into cells, labelled by an index $i$, with size $\epsilon$, which are occupied with probability $p_i$:
\begin{equation}
\label{eq: 2.3}
D_q=\frac{1}{q-1}\lim_{\epsilon\to 0}\frac{{\rm ln}\sum_i p_i^q}{{\rm ln}\,\epsilon}\ .
\end{equation}
The dimension $D_2$ is known as the correlation dimension and is related to the exponent $\beta$ in (\ref{eq: 2.2}). It follows from (\ref{eq: 2.3}) that the number ${\cal N}$ of trajectories in a ball of radius $\epsilon$ about a randomly selected test trajectory satisfies $\langle {\cal N}(\epsilon)\rangle \sim \epsilon^{D_2}$ \cite{Ott02}. The probability density for trajectory separations in one dimension is $\rho(\epsilon)={\rm d}\langle {\cal N}\rangle/{\rm d}\epsilon$, so that $D_2=1-\beta$.

{\sl 3. Analysis of clustering}. We consider the dynamics of a one-dimensional map, of the form
\begin{equation}
\label{eq: 3.1}
x_{n+1}=f_n(x_n)
\end{equation}
where the functions $f_n(x)$ may be independent of $n$ (that is, the map may be deterministic), or they are selected at random from an ensemble at each iteration. We concentrate upon chaotic maps, where the Lyapunov exponent is positive. We will assume that the iterations do not escape from some finite region, so that there is an upper bound on the separation of two trajectories. We allow for the possibility of a random map because the analysis can be taken further in that case and because random maps play a role in modelling many different physical processes.

Both the Lyapunov exponent $\lambda$ and the clustering exponent $\beta$ characterise small separations of trajectories, $\Delta x$. As $\Delta x\to 0$ these become infinitesimal separations, denoted by $\delta x$. The evolution of the infinitesimal separations is described by the linearised equation of motion, $\delta x_{n+1}= f'_n(x_n)\delta x_n$. It is convenient to consider the variable $Y_n={\rm ln}\Delta x_n$, which we express in terms of its incremental changes at each iteration, $Z_n$:
\begin{equation}
\label{eq: 3.2}
Y_n={\rm ln}(\Delta x_n)={\rm ln}(\Delta x_0)+\sum_{i=0}^{n-1} Z_i
\end{equation}
where, in the limit where $\Delta x_i$ is sufficiently small, we have
\begin{equation}
\label{eq: 3.3}
Z_n={\rm ln}\vert f_n'(x_n)\vert\ .
\end{equation}
Even for a deterministic map, the quantity $Z_n$ differs from one iteration to the next, and if the map is chaotic this variation will appear to be random.

We now consider how both the Lyapunov exponent $\lambda$ and the clustering exponent $\beta$ are related to the statistics of $Z$. It is clear from the definitions (\ref{eq: 2.1}) and (\ref{eq: 3.2}) that
\begin{equation}
\label{eq: 3.4}
\lambda=\langle Z(t)\rangle\ .
\end{equation}
Let us consider dynamics of $Y_n$. Because the displacements $Z_n$ can be considered as random variables, the quantity $Y_n$ executes a random walk with drift when $\Delta x_n$ is sufficiently small. The variable $Y$ has diffusive fluctuations, with the variance of $\Delta Y_n=Y_{i+n}-Y_i$ increasing linearly as a function of the \lq time', $n$:
\begin{equation}
\label{eq: 3.5}
\lim_{n\to \infty}\frac{\langle (\Delta Y_n-n \lambda )^2\rangle}{n}=2{\cal D}\ .
\end{equation}
The diffusion coefficient ${\cal D}$ is determined from the correlation function of the fluctuations
\begin{equation}
\label{eq: 3.6}
{\cal D}=\tfrac{1}{2}\sum_{n=-\infty}^\infty \left[\langle Z_n Z_0 \rangle-\langle Z\rangle^2\right]\ .
\end{equation}
We characterise the behaviour of $Y_n$ in terms of its probability density, $\rho_n(Y)$. If the probability density $\rho_n(Y)$ varies sufficiently slowly as a a function of $Y$, we may approximate its evolution by discrete-time version of the Fokker-Planck equation \cite{vKa81}:
\begin{equation}
\label{eq: 3.7}
\rho_{n+1}-\rho_n=-\frac{\partial}{\partial Y}(v\rho_n)+\frac{\partial^2}{\partial Y^2}({\cal D}\rho_n)
\end{equation}
where the drift velocity $v=\langle Z\rangle$ is, by (\ref{eq: 3.4}), equal to the Lyapunov exponent. The steady-state solution of equation (\ref{eq: 3.7}), $\rho_{\rm s}$, is an exponential function:
\begin{equation}
\label{eq: 3.8}
\rho_{\rm s}(Y)=A\exp(\alpha Y)
\end{equation}
where $\alpha =v/{\cal D}$. This solution is not normalisable and we should therefore consider the conditions under which it is applicable. The linearised mapping ceases to be applicable when $\Delta x$ is too large. Because the growth of $\Delta x$ is assumed to be bounded, the probability density $\rho_{\rm s}(Y)$ should have a normalisable steady state. We can therefore meaningfully consider a solution of the form (\ref{eq: 3.8}) with positive values of $\alpha$, because the solution is matched to another function at some cutoff value. However there is
no lower cutoff, so that solutions with negative values of $\alpha$ are untenable.

Now consider the implications of (\ref{eq: 3.8}) for the probability density of trajectory separations. Transforming (\ref{eq: 3.8}) to a density $\rho(\Delta x)$ by writing ${\rm d}P=A\exp(\alpha Y){\rm d}Y=\rho(\Delta x){\rm d}\Delta x$, we find $\rho(\Delta x)=A \Delta x^{\alpha -1}$. We conclude that clustering should occur when $\alpha < 1$, and that the correlation dimension is the same as the exponent in (\ref{eq: 3.8}): we have
\begin{equation}
\label{eq: 3.9}
D_2=\alpha =1-\beta\ .
\end{equation}

The use of the Fokker-Planck equation is only justified when the gradient of $\rho_n(Y)$ is sufficiently small. The condition is that $\partial \rho_n/\partial Y$ should be small compared to $1/\delta Y_0$, where $\delta Y_0$ is the scale over which $Y$ varies during its correlation time. This condition for the validity
of (\ref{eq: 3.8}) is equivalent to $D_2\ll 1$. We remark that Grassberger and Procaccia \cite{Gra+84} also used a Fokker-Planck equation for the logarithm of trajectory separations in a discussion of the correlation dimension. They considered a strange attractor, rather than a system with only positive Lyapunov exponents.

From these considerations we make the following conclusions. For any chaotic map, we can calculate the mean value and variance of $Z_n={\rm ln}|f'_n(x_n)|$. The mean value is equal to the Lyapunov exponent, $\lambda=\langle Z\rangle$, and for a chaotic map this is a positive number. The fluctuations of $Y={\rm ln}\Delta x$ execute a random walk, with diffusion coefficient ${\cal D}$. When the probability density of $Y$ varies sufficiently slowly, this obeys a Fokker-Planck equation. The stationary solution of this equation is an exponential which in turn implies that the distribution of $\Delta x$ is a power-law. The assumption that the probability density is slowly varying is justified when  $D_2=\alpha \ll 1$, a limit in which there is strong clustering. In this limit we have seen that the correlation dimension is given by the asymptotic expression
\begin{equation}
\label{eq: 3.10}
D_2=\lambda/{\cal D}\ .
\end{equation}
Thus clustering is expected to occur for a chaotic one-dimensional map upon varying a parameter so that the Lyapunov exponent approaches zero (from above). The only way to avoid this conclusion is if the diffusion constant ${\cal D}$ approaches zero at the same time as the Lyapunov exponent $\lambda$.

{\sl 4. Correlated random walk model}. We shall now illustrate the clustering phenomenon using a simple random dynamical system, namely a random walk for which the random displacement is a smoothly varying function of the current position. The map is
\begin{equation}
\label{eq: 4.1}
x_{n+1}=x_n+g_n(x_n)
\end{equation}
and the function $g_n(x)$ is a realisation of a random process with statistics
specified by a function $C(x-x')$:
\begin{equation}
\label{eq: 4.2}
\langle g_n(x)\rangle=0
\ ,\ \ \
\langle g_n(x)g_m(x')\rangle=\delta_{nm}C(x-x')\ .
\end{equation}
Furthermore we consider the coordinate $x$ to be cyclic, so that the positions $x$ and $x+L$ are equivalent. (This implies that the functions $g_n(x)$ are periodic with period $L$). This model has been discussed in earlier work \cite{Wil+03} where it was pointed out that the Lyapunov exponent is negative when the functions $g_n(x)$ are sufficiently small. The model is of some physical interest as a model for the motion of particles advected by a spatially smooth velocity field, such as a fluid flow \cite{Wil+03}. Its attraction in the present context is that it allows us to write down and analyse an exact (although implicit) equation for the correlation dimension, $D_2=\alpha$.

Because there is no correlation between the random displacements at successive iterations, the probability density $\rho_{n+1}$ for $Y_{n+1}$ may be expressed exactly in terms of density $\rho_n$ of $Y_n$. For sufficiently small $\Delta x$ we have:
\begin{equation}
\label{eq: 4.3}
\rho_{n+1}(Y)=\int_{-\infty}^\infty {\rm d}Z\ P(Z)\rho_n(Y-Z)
\end{equation}
where $P(Z)$ is the probability density of $Z_n=|1+g'(x_n)|$. This equation has a steady-state solution of the form (\ref{eq: 3.8}), that is $\rho_{\rm s}(Y)=A \exp(\alpha Y)$ for some constant $A$ and for an appropriate choice of $\alpha$. By substituting (\ref{eq: 3.8}) into (\ref{eq: 4.3}), we find that $\alpha$ satisfies
\begin{equation}
\label{eq: 4.4}
\langle \exp(-\alpha Z)\rangle=\int_{-\infty}^\infty {\rm d}Z\ \exp(-\alpha Z)P(Z)=1
\ .
\end{equation}
It is a simple exercise to reproduce equation (\ref{eq: 3.10}) from this equation by writing $\langle\exp(-\alpha Z)\rangle=1-\alpha \langle Z\rangle+\frac{1}{2}\alpha^2[\langle Z^2\rangle-\langle Z\rangle^2]+O(\alpha^3)$. If equation (\ref{eq: 4.4}) has a solution with $0<\alpha<1$, this demonstrates
conclusively that fractal clustering exists for a model with only positive Lyapunov exponents.

\begin{figure}[t]
\centerline{\includegraphics[width=9.0cm]{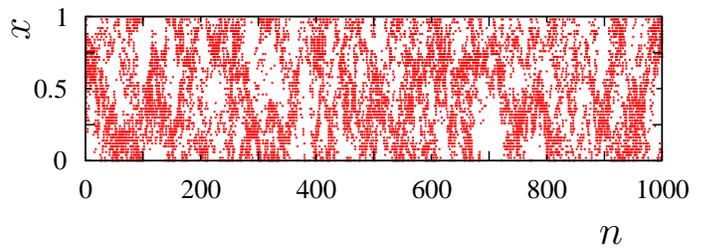}}
\caption{\label{fig: 1} Illustrating clustering of the trajectories of the correlated random walk model defined by equations (\ref{eq: 4.1}) and (\ref{eq: 4.2}).
}
\end{figure}

We made a numerical investigation of the model described by (\ref{eq: 4.1}) and (\ref{eq: 4.2}), in the case where the noise has a Gaussian distribution and where the correlation function is $C(x)=\sigma^2\xi^2\exp(-x^2/2\xi^2)$ (for some constants $\sigma$ and $\xi$). The density of trajectories in this model is illustrated in figure \ref{fig: 1}, where we show a scatter plot of the positions for a realisation of the dynamics described by (\ref{eq: 4.1}) and (\ref{eq: 4.2}) as a function of iteration number, for a case where the Lyapunov exponent is positive. This shows that the trajectories cluster together, rather than becoming uniformly distributed, as might be expected. The probability density $\rho(\Delta x)$ for a trajectory to occur at a distance $\Delta x$ from a randomly chosen test trajectory was found to have a power-law form, as described by equation (\ref{eq: 2.2}), implying that the distribution can be regarded as a fractal set, with correlation dimension $D_2$.

\begin{figure}[t]
\centerline{\includegraphics[width=6.5cm]{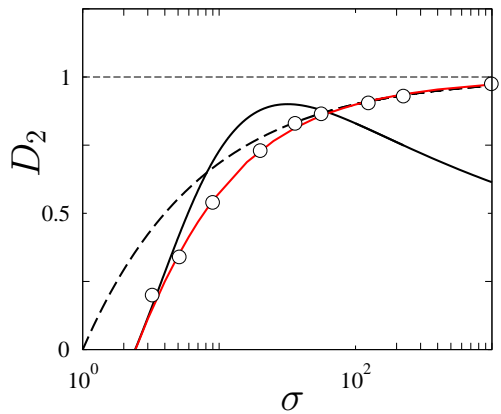}}
\caption{\label{fig: 2} The correlation dimension of the random map (\ref{eq: 4.1}), (\ref{eq: 4.2}), determined by solving (\ref{eq: 4.5}) (solid line, red in colour) is compared with exponents determined by simulation ($\circ$). The asymptotic approximations to $D_2$, (\ref{eq: 3.10}) (solid black line) and (\ref{eq: 4.6}) (dotted line) are also shown.
}
\end{figure}

As well as confirming that clustering occurs, it is interesting to consider the quantitative predictions from (\ref{eq: 4.4}) in the case where the random functions $g_n(x)$ have Gaussian statistics. Specifically, we assume that the quantities $G_n=g'(x_n)$ are Gaussian distributed, with variance $\sigma^2$ (the mean value is obviously zero). In this case the relation (\ref{eq: 4.4}) becomes
\begin{eqnarray}
\label{eq: 4.5}
1&=&\int_{-\infty}^\infty {\rm d}G\ \delta (Z-{\rm ln}\vert 1+G\vert) P(G) \exp(-\alpha Z)
\nonumber \\
&=&\frac{1}{\sqrt{2\pi}\sigma}\int_{-\infty}^\infty {\rm d}G\ \exp(-G^2/2\sigma^2)
\vert 1+G\vert^{-\alpha}
\nonumber \\
&=&\frac{1}{\sqrt{\pi}2^{\alpha/2}\sigma^\alpha}\Gamma\left(\frac{1-\alpha}{2}\right){}_1F_1\left[\frac{\alpha}{2};\frac{1}{2};-\frac{1}{2\sigma^2}\right]
\ .
\end{eqnarray}
where $\Gamma$ is the Euler gamma function and ${}_1F_1$ is the Kummer confluent hypergeometric function. In the limit as $\sigma \to \infty$ the exponent $D_2=\alpha $ approaches unity as
\begin{equation}
\label{eq: 4.6}
D_2 \sim 1-\sigma^{-1}
\end{equation}
so that the clustering tendency is always present in (\ref{eq: 4.1}), (\ref{eq: 4.2}), no matter how strong the random impulses $g_n(x)$. The value of $\alpha$ which solves equation (\ref{eq: 4.5}) is plotted as a function of $\sigma^2$ in figure \ref{fig: 2}, where it is compared with the asymptotic approximations, (\ref{eq: 3.10}) and (\ref{eq: 4.6}) and with values of the exponent obtained by simulation.

5. {\sl Extension to deterministic maps}. One dimensional deterministic maps can be chaotic, with a positive Lyapunov exponent. It is natural to ask whether these can also exhibit clustering behaviour. The best studied example \cite{Ott02} is the logistic map, $x_{n+1}=Kx_n(1-x_n)=f(x_n)$ which has been studied exhaustively and which is considered to have generic properties. There are intervals of the parameter $K$ where the trajectory is attracted to stable periodic orbits, interspersed with sets where the Lyapunov exponent is positive. When the correlation dimension is small it is well approximated by $D_2=\lambda/{\cal D}$. We therefore expect that clustering of trajectories will occur close to the boundaries of the chaotic regions, where $\lambda$ is small.

A numerical investigation of the correlation dimension for the logistic map was carried out, for a large set of values of the coupling parameter $K$. This set included many values close to the edge of a stable region, where the Lyapunov exponent, although positive, was very small. This investigation produced a surprising result. With one exception, no values of $K$ were identified for which the numerically determined correlation dimension was clearly different from unity (in the chaotic zones) or zero (in the stable zones). The only exception was when $K$ was set equal to the value which corresponds to the limit point of the Feigenbaum period-doubling sequence. At that value of $K$, the results suggested a value of $D_2$ which is less than unity but greater than zero, consistent with results by Grassberger \cite{Gra81}.

The observation that $D_2$ does not become small when $\lambda$ is small is consistent with (\ref{eq: 3.10}) if the diffusion constant ${\cal D}$ approaches zero at the same time as $\lambda$ approaches zero. In the case of deterministic systems, this is in fact what happens, as the following heuristic argument shows.
Consider what happens as the parameter $K$ of the logistic map is varied from $K_-$, where $\lambda$ is slightly negative, to the nearby value $K_+$ where $\lambda$ is slightly positive. At $K_-$ the map has an attractor which is a stable periodic orbit of period $M$ (say). The Lyapunov exponent for the stable orbit is
\begin{equation}
\label{eq: 5.4}
\lambda=\frac{1}{M}\sum_{j=1}^M {\rm ln}|f'(x_j^\ast)|
\end{equation}
where $x_j^\ast$ are the points on the periodic orbit. In the case of the periodic orbit the shift of $Y$ with every orbit is precisely $M\lambda$, implying that ${\cal D}=0$ for the stable orbit. At $K_+$ the periodic orbit has either become unstable, or else has ceased to exist. The trajectory then explores a dense subset of the line, but it must spend most of the time in the vicinity of the sequence of $M$ points visited by the nearby stable orbit. While the trajectory is trapped close to the periodic orbit, there is negligible randomness in the values of $Z$. We therefore expect that ${\cal D}$ approaches zero at the boundary of the unstable region. If a small amount of noise is added to a deterministic map, the value of ${\cal D}$ remains finite at the boundary of the chaotic region, and we find that $D_2$ approaches zero smoothly at the boundary.

{\sl Conclusions}. Fractal clustering may occur even if there is no negative Lyapunov exponent. This was analysed and demonstrated in a random dynamical system, namely the correlated random walk. In the case of the logistic map, the clustering effect was not observed. We argued that it is harder to see the clustering effect in a deterministic dynamical system, although there does not appear to be any reason why it is forbidden. It would be interesting to find an example.

{\em Acknowledgments.}
The work of BM was supported by Vetenskapsr\aa{}det, and that of BM, KG and EW by the platform \lq Nanoparticles in an interactive environment' at G\"oteborg university.


\begin{thebibliography}{}

%
\bibitem{Eck+85}
J.-P. Eckmann and D. Ruelle,
{\it Rev. Mod. Phys.}, {\bf 57}, 617–656, (1985).
%
\bibitem{Ott02}
E. Ott,
{\sl Chaos in Dynamical Systems}, 2nd edition, Cambridge: University Press, (2002).
%
\bibitem{Hen+83}
H. G. E. Hentschel and I. Procaccia,
{\it Physica D}, {\bf 8}, 435, (1983).
%
\bibitem{Kap+79}
J. L. Kaplan and J. A. Yorke, in {\sl Functional Differential Equations and
Approximations of Fixed Points}, Lecture Notes in Mathematics, eds.
H.-O. Peitgen and H.-O. Walter, Springer, Berlin, {\bf 730}, 204, (1979).
%
\bibitem{Led+88}
F. Ledrappier and L. S. Young,
{\it Comm. Math. Phys.} {\bf 117}, 529, (1988).
%
\bibitem{vKa81} N. G. van Kampen,
{\em Stochastic processes in physics and chemistry}, 2nd ed.,
North-Holland, Amsterdam, (1981).
%
\bibitem{Gra+84}
P. Grassberger and I. Procaccia,
{\it Physica D}, {\bf 13}, 34-54, (1984).
%
\bibitem{Wil+03}
M. Wilkinson and B. Mehlig,
{\it Phys. Rev. E}, {\bf 68}, 040101(R), (2003).
%
\bibitem{Gra81}
P. Grassberger, {\it J. Stat. Phys.}, {\bf 26}, 173-9, (1981).
%
\end{thebibliography}
\end{document}